\documentclass[a4paper]{jpconf}
\usepackage{graphicx}
 \usepackage{amssymb}
\usepackage{cite}

\newcommand{\nc}[1]{\newcommand{#1}}

\nc{\its}[1]{\itshape #1 \upshape}
\nc{\mc}[3]{\multicolumn{#1}{#2}{#3}}

\nc{\bc}{\begin{center}}
\nc{\ec}{\end{center}}
\nc{\ig}[1]{\bc \includegraphics{#1} \ec}

\nc{\bo}[1]{\mbox{\boldmath \( #1 \! \! \)  \unboldmath}}
\nc{\be}{\begin{eqnarray}}
\nc{\ee}{\end{eqnarray}}
\nc{\bew}{\begin{eqnarray*}}
\nc{\eew}{\end{eqnarray*}}

\begin{document}
\title{Exploring QCD phase diagram at vanishing baryon density on the lattice }

\author{Heng-Tong Ding}
\address{Physics Department, Brookhaven National Laboratory, Upton NY 11973 }

\ead{htding@quark.phy.bnl.gov}

\begin{abstract}

I report on the current status of QCD phase diagram at vanishing baryon density.  I focus on the QCD phase diagram with three 
degenerate quark flavor using Highly Improved Staggered Quarks on $N_{\tau}=6$ lattices. No evidence of a first order phase transition in the pion mass window of $80\lesssim  m_{\pi} \lesssim 230~$MeV is found.
The pion mass at the critical point where the chiral first order phase transition ends is estimated to be $m^c_{\pi} \lesssim 45$ MeV.
\end{abstract}

\section{Introduction}

\begin{figure}[htp]
\begin{center}
\includegraphics[width=0.3\textwidth]{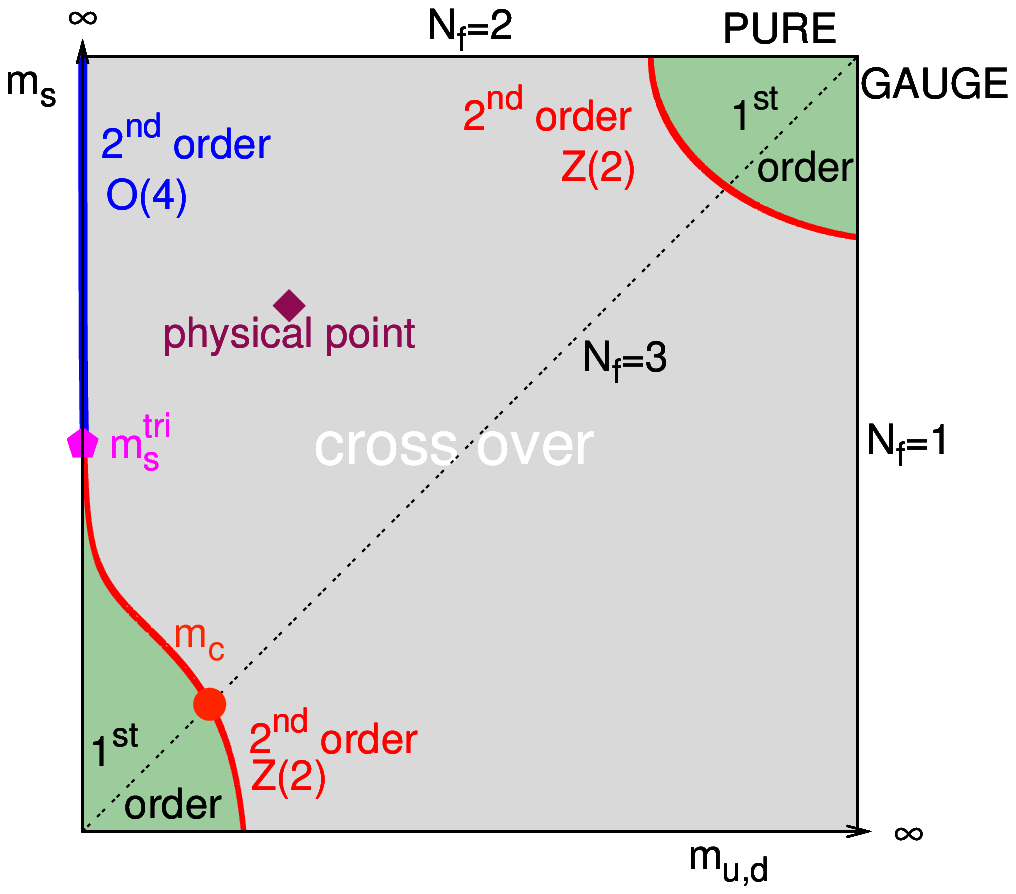}~~~
\includegraphics[width=0.4\textwidth,height=0.3\textwidth]{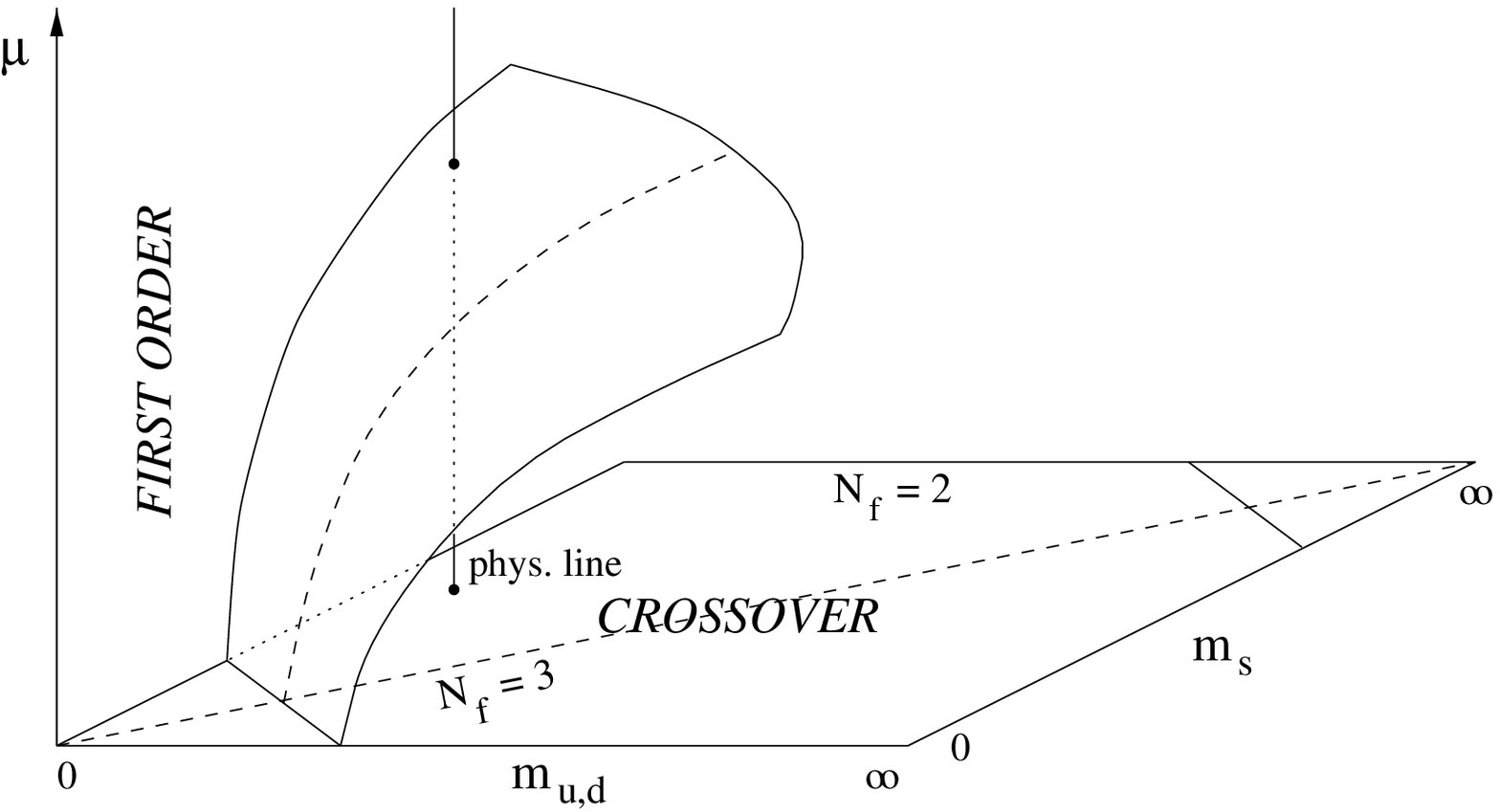}~
\end{center}
\caption{Left: schematic QCD phase transition behavior for different choices of quark 
masses ($m_{u,d}$, $m_{s}$) at zero chemical potential. Right: The critical 
surface swept by the chiral critical line at finite chemical potential. A QCD chiral critical point may exist if the surface bends towards the physical point. The left and right 
plots are taken from Ref.~\cite{HISQmc} and Ref.~\cite{Karsch04}, respectively.}
\label{fig:sketch}
\end{figure}

Due to the asymptotic freedom feature of QCD, it is expected that the QCD matter will undergo a phase transition from
hadronic phase to the quark gluon plasma phase at sufficient high energy density. The nature of such a phase transition crucially depends on the values of quark 
masses (light $m_{l}$ and strange $m_s$), the number of quark flavors ($N_f$) and the baryon density ($\mu_B$)~\cite{KarschLecture02, reviews}~\footnote{We discuss implicitly the case with number of colors $N_c=3$  in this proceedings.}. 
The conjectured QCD phase diagram at vanishing baryon density from effective theories on the basis of the renormalized group together with the universality~\cite{conj} is summarized in the left plot of Fig.~\ref{fig:sketch}.
As sketched in Fig.~\ref{fig:sketch}, in the pure gauge case, i.e. in the limit of $m_l\rightarrow \infty$
and $m_s\rightarrow \infty$, QCD has an exact Z(3) symmetry and the phase transition is of first order. In the chiral limit of $N_f=3$ case, i.e. $m_l=m_s\rightarrow 0$, the phase transition
is also known to be a first order. In the chiral limit of a $N_f=2$ theory, i.e. $m_l\rightarrow 0$ and $m_s\rightarrow \infty$, if $U_A(1)$ symmetry is restored the relevant symmetry
is $O(2)\times O(4)$ and the transition is likely first order\footnote{A second order phase transition is also allowed with a different symmetry breaking pattern~\cite{Basile:2005hw}.}; if $U_A(1)$ symmetry is broken the relevant symmetry becomes isomorphic to that of the 3-d O(4) spin model
and the transition becomes second order belonging to that universality class. In the intermediate quark mass region, there is no real phase transition but a (rapid) crossover takes place from the hadronic phase to the quark gluon
plasma phase.
All the first order phase transition regions are separated from the crossover by lines of second order phase transitions belonging to the 3-d Ising $Z(2)$ universality class.
Thus the first order region for $N_f=3$ case, the second order O(4) line for $N_f=2$ case and
the second order $Z(2)$ line are supposed to meet at a tri-critical point characterized by a certain value $m_s^{tri}$ of the strange quark mass. For the system with massless $N_f\geq 3$ flavors, the order of its chiral phase transition is always first order and is independent on the fate of $U_A(1)$ symmetry.

The QCD phase diagram at vanishing baryon density has been studied extensively on the lattice.
It has been well established that QCD with $N_f=0$ has a first order phase transition from a confined phase to a deconfined phase at the transition temperature $T_c\approx 270$ MeV~\cite{Karsch:1999vy}. 
With physical values of the quark masses
$m_u=m_d=m_l^{phys}$ and $m_s=m_s^{phys}$ it is found from lattice simulations with staggered fermions that QCD does not have a true transition but a rapid crossover
and the pseudo critical transition temperature $T_{pc}\approx 154$ MeV~\cite{Tpc}. 
The location of the physical point with respect to the tri-critical point is not fully determined. More specifically, it is not yet clear whether
$m_s^{phys}>m_s^{tri}$ or $m_s^{phys}=m_s^{tri}$ or $m_s^{phys}<m_s^{tri}$. If $m_s^{phys}>m_s^{tri}$ then
in the limit of $m_l\to0$ one should see a second order transition belonging to the
$3$-d $O(4)$ universality class, if $m_s^{phys}=m_s^{tri}$ the tri-critical point is a Gaussian fixed point of the 3-dimensional $\phi^6$ model and its critical 
exponents take the mean field values~\cite{Riedel72} and if $m_s^{phys}<m_s^{tri}$ then
in $m_l\to0$ limit first one should cross through a second order transition belonging
to the $3$-d $Z(2)$ universality class and then end up in the first order transition
region. Recent lattice QCD calculations with improved staggered fermions indicate that QCD transition in the chiral limit of $m_l$ with $m_s=m_s^{phys}$ is likely 
of second order and belongs to O(4) (O(2)) universality class~\cite{Ejiri:2009ac}. To confirm this scenario, further study with better discretization scheme of QCD action
and the investigation on the fate of $U_A(1)$ symmetry at finite temperature~\cite{Ding:2012ar} are crucially needed.

For QCD with $N_f\geq 3$ a first order phase transition has been observed at small quark mass on coarse lattices using unimproved~\cite{1stOrder} as well as the p4-improved staggered fermions~\cite{1stOrder_p4}.
However, the value of pion mass at the critical point where the first order phase transition ends, $m_\pi^c$, is not yet determined in the continuum limit and its value at finite lattice cutoff varies from $\sim$300 MeV to $\sim$50 MeV, strongly depending on the lattice spacing and discretization schemes of the action~\cite{Karsch04,1stOrder,stout,HISQmc}.
The value of $m_\pi^c$ and the curvature of the surface swept from the chiral first order phase transition region at finite $\mu_B$ may be relevant to the (non-)existence of a QCD chiral critical point as sketched in the right plot
of Fig.~\ref{fig:sketch} although other possibilities for generating a second order transition at the physical values of quark masses have been discussed in Ref.~\cite{Gupta:2008ac}.
Recent lattice QCD studies based on the imaginary chemical potential approach disfavor the existence of such a QCD chiral critical point~\cite{deForcrand:2008vr}. This result, however, is obtained from simulations using unimproved staggered fermions on coarse lattices and may be prominently influenced by lattice cutoff effects. Thus it would be interesting to investigate this topic using an improved fermion action on finer lattices. As a starting point to determine the curvature of the chiral critical surface, the value of $m_\pi^c$ needs to be determined. In this talk
we thus focus on the QCD phase diagram with three degenerate Highly Improved Staggered Quarks at vanishing baryon density and discuss the value of $m_\pi^c$.

\section{Universality class near critical lines}

As mentioned above in the chiral limit of three flavor QCD the transition is of first order. This first order phase transition becomes weaker with finite quark mass and ends
at a critical quark mass $m_c$ sitting on a second order phase transition line. The universal properties of this critical point belong to a $Z(2)$ symmetry. 
The proper order parameter could be a mixture of two relevant quantities, e.g. a combination of the chiral condensate with the pure gauge action $S_G$~\cite{HISQmc}
%\be
$, M= ( \langle\bar{\psi} \psi \rangle + r S_G) \Big{|}_{\rm T=T_c}   \propto  (m-m_c)^{1/\delta}$
%\ee
and the susceptibility of the order parameter M, i.e.
%\be
$\chi_{M}/T^2  \Big{|}_{\rm T=T_c}  \propto  (m-m_c)^{1/\delta-1}$.
%\label{eq:chi_scaling}
%\ee
Here for $Z(2)$ universal class $1/\delta-1$=-0.785. To estimate the value of $m_c$, we ignore the contribution from $S_G$ and will use the quark chiral condensate as an approximate order parameter for the second order phase transition at the critical point.

\begin{table}[hptb]
\begin{center}
\small
\begin{tabular}{|c|c|c|c|c|}
\hline
lattice dim. &   $am_q$     & $m_{\pi}$ [MeV]     &\# $\beta$ values &  max no. of traj. \\
\hline
$16^3\times$ 6           & 0.0075          & 230                       &  17                  & 8990             \\
$10^3\times$ 6          & 0.00375        & 160                    &  12                    & 12900             \\
$12^3\times$ 6          & 0.00375        & 160                    &  12                    & 12900             \\
$16^3\times$ 6          & 0.00375        & 160                    &  12                    & 12900             \\
$24^3\times$ 6          & 0.00375        & 160                    &  12                    & 12690             \\
$24^3\times$ 6          & 0.0025        &  130                       &   5                     & 11900           \\
$24^3\times$ 6           & 0.001875      & 110                      &  7                    & 8990           \\
$24^3\times$ 6          & 0.00125        &  90                       &   7                     & 10190           \\
$24^3\times$ 6           & 0.0009375    & 80                       &   8                     & 11790           \\
$16^3\times$ 6        & 0.0009375    &  80                   &   6                    & 7670          \\
\hline
\end{tabular}
\end{center}
\caption{
Parameters of the numerical simulations.
}
\end{table}

As mentioned in Ref.~\cite{HISQmc} the chiral first order phase transition region from lattice QCD simulations shrinks with reduced cutoff 
effects. This is also observed by the lattice QCD calculation using the stout action~\cite{stout}. The Highly Improved Staggered Quark (HISQ) action  
achieves better taste symmetry than the 
asqtad and p4 actions at a given lattice spacing~\cite{HISQ}. 
Simulations presented here have been carried out with three degenerate quark flavors using the HISQ action. The quark masses vary from 0.0075 to 0.0009375 corresponding to pion masses in the region of $80 \lesssim m_{\pi} \lesssim 230~$MeV. 
In this proceeding we update the results reported in Ref.~\cite{HISQmc} by performing additional simulations with $am=0.00375$ on $16^3\times 6$, $12^3\times 6$ and $10^3\times 6$ lattices and simulations with quark mass $am=0.025$ on $24^3\times6$ lattices.

\section{Results}

\begin{figure}[htp!]
\begin{center}
\includegraphics[width=.32\textwidth]{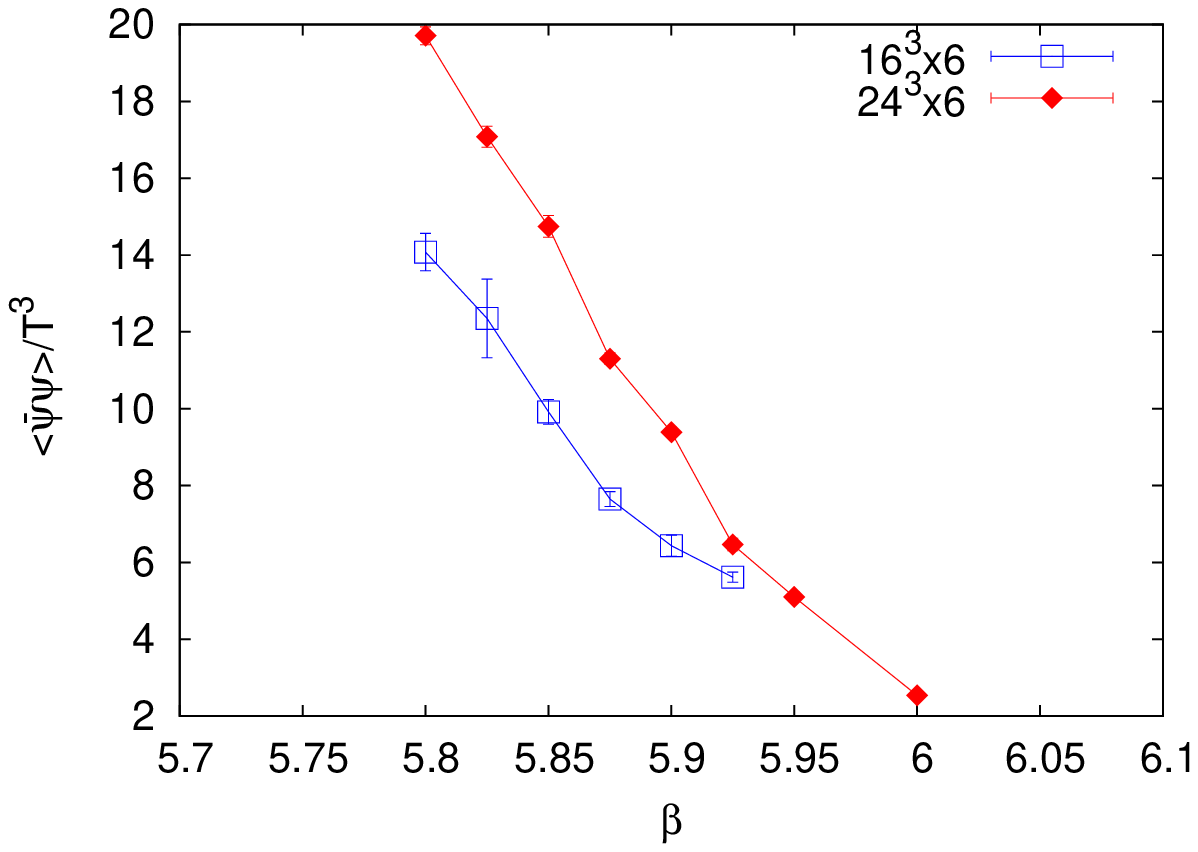}~~~\includegraphics[width=.32\textwidth]{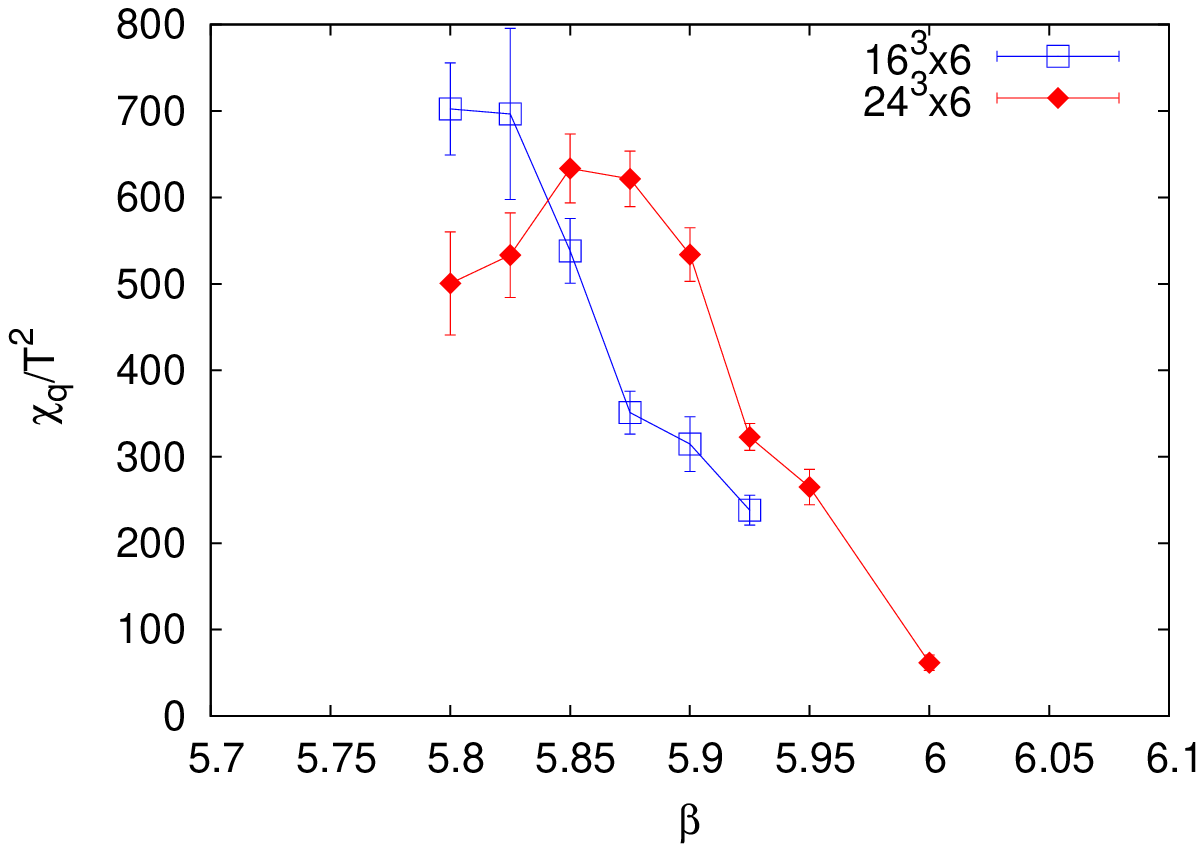}
\end{center}
\caption{Volume dependences of  chiral condensates (left) and chiral susceptibilities (right) with quark mass $am=0.0009375$.}
\label{fig:volume_dep}
\end{figure}

We show the volume dependence of quark chiral condensate with $am=0.0009375$ ($m_{\pi}\approx 80 $ MeV) in the left plot of Fig.~\ref{fig:volume_dep}.
The volume dependence at high temperature is small while at low temperature it is relatively large. This is obvious since
at high temperature the scale of the system is controlled by the temperature $T$ while at low temperature the scales are the chiral symmetry breaking scales.
In the right plot of Fig.~\ref{fig:volume_dep} we show the volume dependence of chiral susceptibility. If the system with quark mass $am=0.0009375$
is in the first order phase transition region, the chiral susceptibility should scale with volume. However, such volume scaling behavior is not observed in the system with pion mass $m_{\pi}\approx80$ MeV.

\begin{figure}[htp!]
\begin{center}
\includegraphics[width=.32\textwidth]{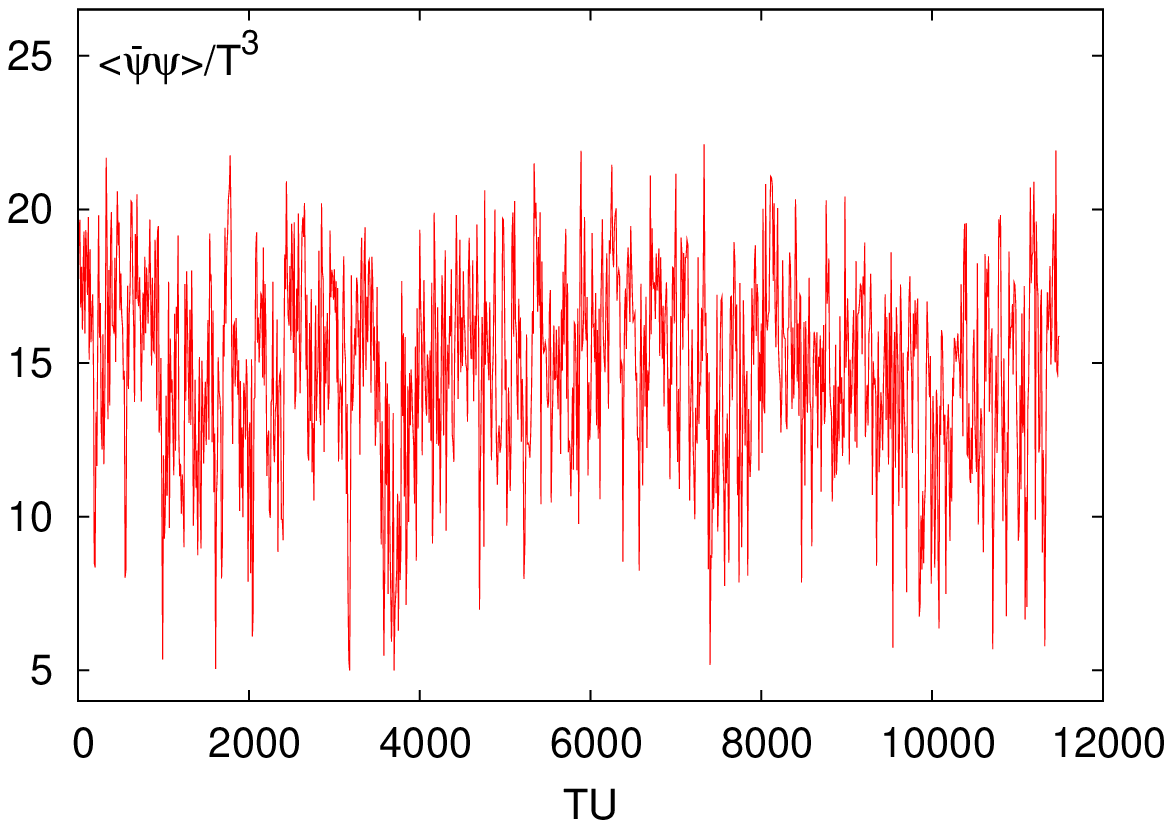}~~~\includegraphics[width=.32\textwidth]{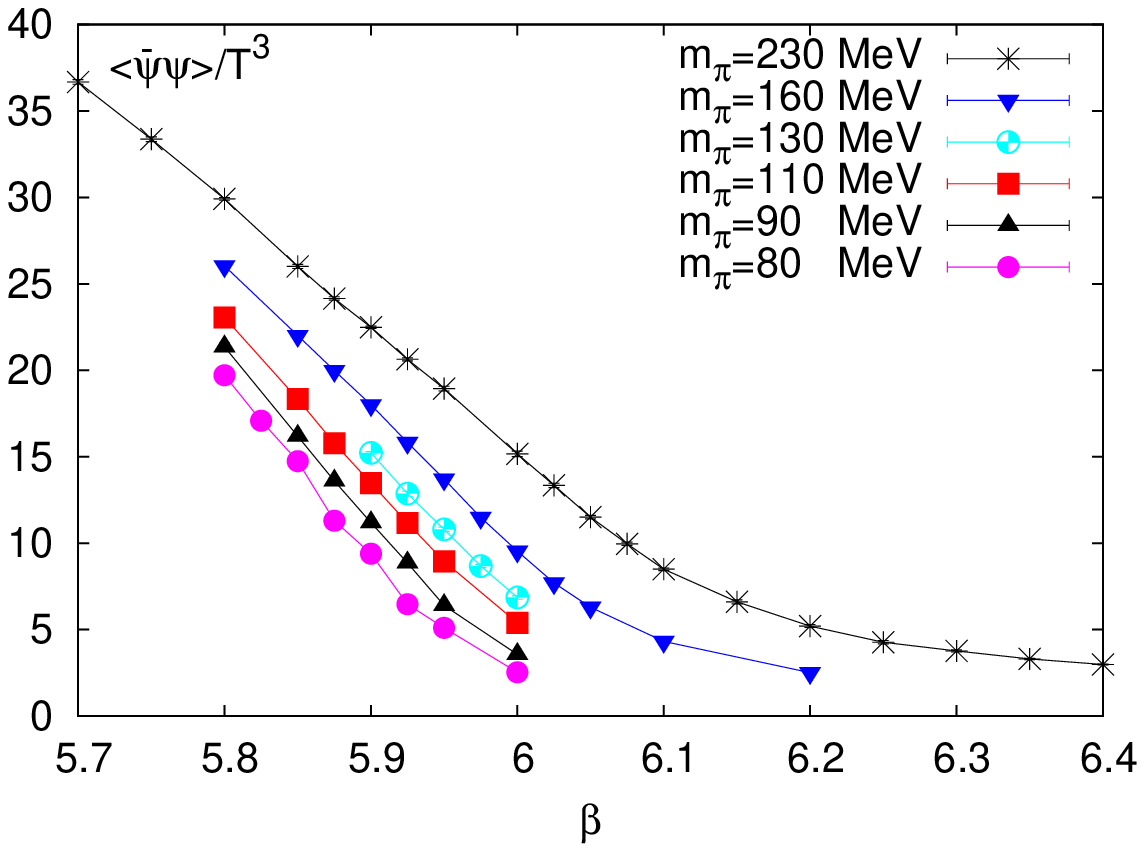}%~~\includegraphics[width=.45\textwidth]{binder_cumulant.eps}
\end{center}
\caption{Left: time history of the quark chiral condensate near $\beta_c$ with $am=0.0009375$ on $24^3\times6$ lattices. Right: Chiral condensates as a function of $\beta$.}
\label{fig:signal}
\end{figure}

In the left plot of Fig.~\ref{fig:signal}, we show the time history of quark chiral condensate near the pseudo critical $\beta$ value for our lowest quark mass on $24^3\times6$ lattices. There is no evidence for the coexistence of two phases.  We then investigate the temperature dependence of the chiral condensate at different quark masses shown in the right plot of Fig.~\ref{fig:signal}. No evidence for a discontinuity in $\langle \bar{\psi}\psi\rangle$ as function of $\beta$ at all quark masses is found. Together with the evidence from Fig.~\ref{fig:volume_dep} we conclude that there is no first order phase transition in the system with pion mass ranging from 230 to 80 MeV.

\begin{figure}[htp!]
\begin{center}
\includegraphics[width=.35\textwidth]{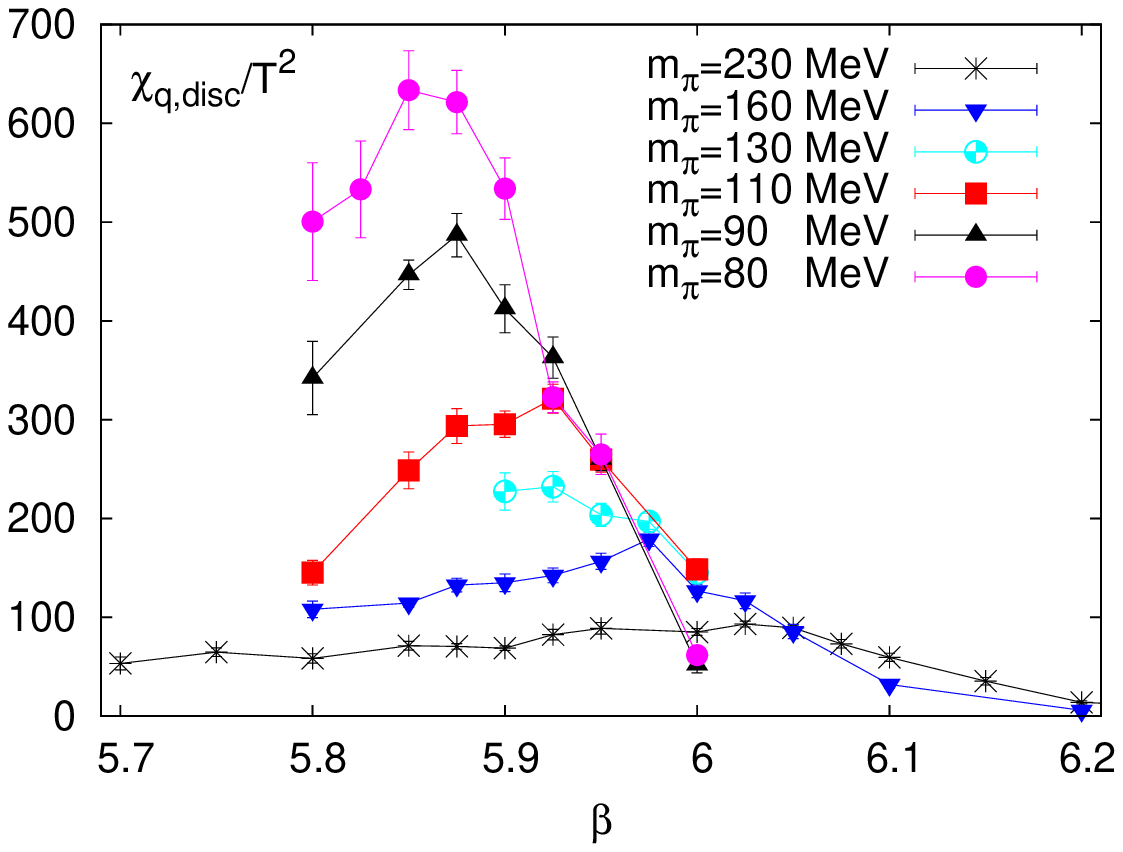}~~~~\includegraphics[width=.35\textwidth]{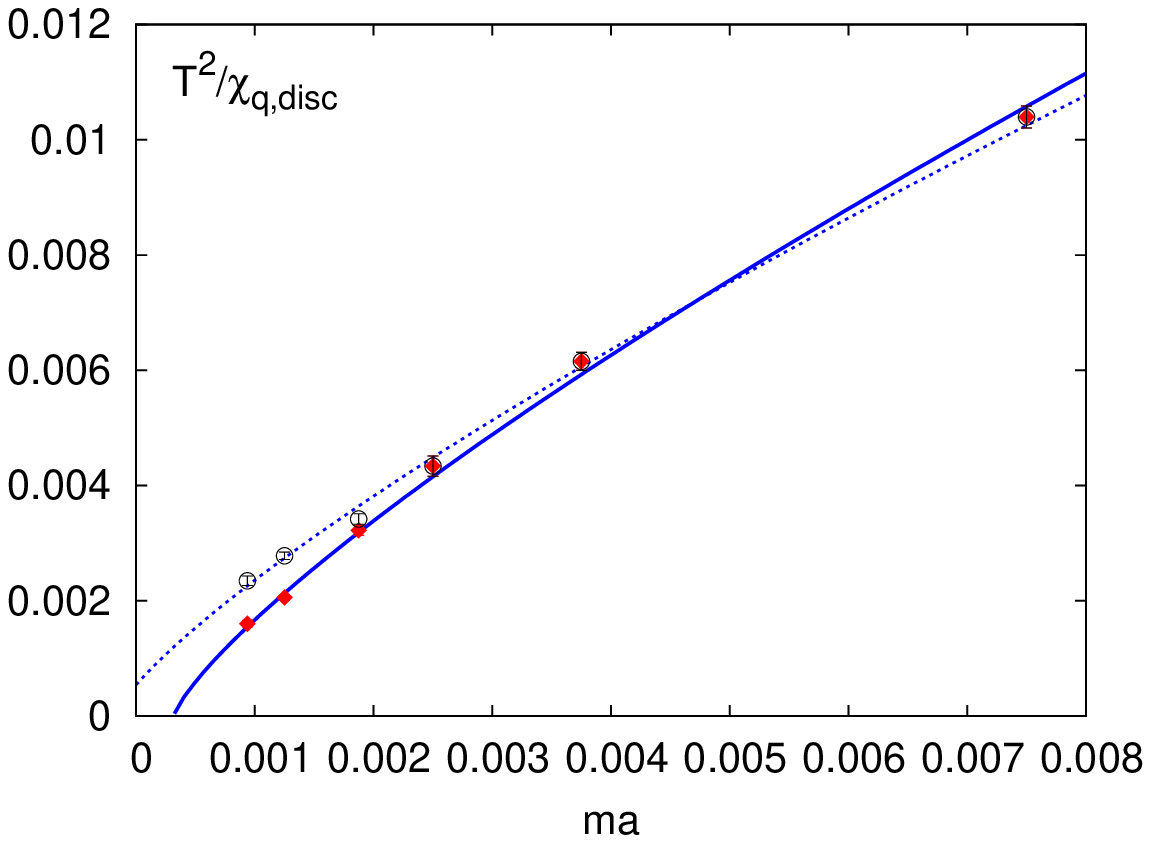}
\end{center}
\caption{Left: the disconnected part of chiral susceptibility as a function of $\beta$. Right: the scaling fit to the inverse height of chiral susceptibility peaks. The filled points denote the inverse peak heights of chiral susceptibilities on largest volume at all quark masses while the open points are rescaled from the filled points by the volume effects. The solid line and dashed line are fitting curves using Z(2) critical exponents.}
\label{fig:chi_s}
\end{figure}

In the left plot of Fig.~\ref{fig:chi_s} we show the disconnected part of the chiral susceptibility as a function of $\beta$ at all quark masses. The pseudo critical temperature decreases with decreasing quark mass. It is expected as the hadronic degrees of freedom in the system become lighter and thus they become more easily excited in the thermal heat bath.  They then can contribute to the energy density of the system and thus trigger the onset of a phase transition already at a lower temperature. The peak height of chiral susceptibility grows with decreasing quark mass as the system is approaching the first order phase transition region. We then performed a scaling fit using an ansatz of $A(m-m_c)^{1-1/\delta}$ to the inverse heights of chiral susceptibility peaks. The results are shown in the right plot of  Fig.~\ref{fig:chi_s}. 
The intercept of the solid line with the $x$ axis gives an estimate for the critical quark mass. Taking into account that in the infinite volume limit the inverse peak height becomes larger (see Fig.~\ref{fig:volume_dep} left) an upper bound for the value of the critical pion mass at the critical point can be obtained, i.e. $m_\pi^{c} \lesssim 45$ MeV.

\section{Conclusion}

We have performed 3-flavor QCD simulations using the HISQ action on $N_{\tau}=6$ lattices with six pion masses in the region of $80\lesssim  m_\pi \lesssim~$230 MeV.
Through the study of quark chiral condensates and chiral susceptibilities, we found no evidence of a first order chiral phase transition in this pion mass window. 
The upper bound on the pion mass at the critical point is estimated to be around $45$ MeV by Z(2) universal scaling analysis considering the finite volume effects.
 In the quark mass plane this upper bound for the critical point in the 3-flavor QCD is thus at about ($m_{\rm phy}^{s}/270,m_{\rm phy}^{s}/270$), which is far away from the physical point at ($m^{s}_{\rm phy}/27,m^{s}_{\rm phy}$). Together with the results from Ref.~\cite{stout}, our results suggest that the first order phase transition region is very small and thus the critical surface swept by the chiral critical line at finite chemical potential has to bend towards the physical point with a very large curvature to affect the nature of the transition at physical values of quark masses at a small chemical potential.

\section{Acknowledgements}

The numerical simulations were carried out on clusters of
the USQCD Collaboration in Jefferson Lab and Fermilab, and on BlueGene computers at the New York Center for Computational Sciences (NYCCS)
at Brookhaven National Lab. This manuscript has been authored under contract number DE-AC02-98CH10886 with 
the U.S. Department of Energy.

\end{document}